\documentclass[a4paper,prd,preprintnumbers,aps,superscriptaddress,nofootinbib,showpacs,preprintstyle]{revtex4-1} 

\usepackage[utf8x]{inputenc}

\usepackage{epsfig}
\usepackage{latexsym}
\usepackage{graphics}
\usepackage{graphicx}
\usepackage{amsmath}
\usepackage{amsfonts}
\usepackage{amssymb}
\usepackage{float}
\usepackage{bm}

\newcommand{\mathsym}[1]{{}}

\newcommand{\baz}{\begin{array}{cc}}
\newcommand{\bad}{\begin{array}{ccc}}
\newcommand{\bi}{\begin{itemize}}
\newcommand{\ei}{\end{itemize}}
\newcommand{\ba}{\begin{array}{c}}
\newcommand{\ea}{\end{array}}

\def\be{\begin{equation}}
\def\ee{\end{equation}}
\def\bea{\begin{eqnarray}}
\def\eea{\end{eqnarray}}
\def\gsim{\ \rlap{\raise 2pt\hbox{$>$}}{\lower 2pt \hbox{$\sim$}}\ }
\def\lsim{\ \rlap{\raise 2pt\hbox{$<$}}{\lower 2pt \hbox{$\sim$}}\ }
\def\dslash{\kern-4pt \not{\hbox{\kern-2pt $\partial$}}}
\def\pslash{\not{\hbox{\kern-2pt p}}}

\def\pmue{{${P_{\mu e}}$ }}

\def\numutonue{{{{\nu_\mu \to \nu_e}} }}

\def\dcp{{\delta_{\mathrm{CP}}}}
\def\pmue{{{ P_{\mu e}}}}
\newcommand{\nova}{NO$\nu$A}

\newcommand{\dm}[1]{\Delta m^2_{#1}}
\newcommand{\ahat}{\hat{A}}

\begin{document}

\DeclareGraphicsExtensions{.eps,.ps}


\renewcommand{\thefootnote}{\fnsymbol{footnote}}

\title{Matter effects at the T2HK and T2HKK experiments}

\author{Sushant K. Raut}
\email{sushant@ibs.re.kr} 
\affiliation{Center for Theoretical Physics of the Universe, Institute for Basic Science (IBS), 
   Daejeon, 34051, South Korea}

\begin{abstract}
Determining the neutrino mass hierarchy and measuring the CP-violating phase $\dcp$ are two of the 
main aims in neutrino physics today. 
The upcoming T2HK (with small matter effects and high statistics) and DUNE (with large matter effects) 
experiments have been shown to have excellent sensitivity to $\dcp$ and the neutrino mass hierarchy, 
respectively. The recent T2HKK proposal aims to improve the hierarchy sensitivity of T2HK by 
placing one of the two tanks of the
HK detector at a site in Korea, to collect data at $\sim 1100$ km baseline. 
In light of the fact that DUNE will anyway collect data at $\sim 1300$ km, we explore whether it 
is advantageous to collect additional long-baseline data as proposed with T2HKK, or to enhance the $\dcp$-precision with the `conventional' 
T2HK by keeping both detector tanks in Japan. We do this by comparing the physics reach of these two options in conjunction with DUNE. 
We find that DUNE+T2HKK is better at excluding the wrong hierarchy, reaching $\sqrt{\Delta\chi^2} \gsim 13$ irrespective 
of the true parameters. While DUNE+T2HK can measure $\dcp$ with 
more precision in some parts of the parameter space, both DUNE+T2HK and DUNE+T2HKK perform equally well near the 
current best-fit point, giving a $\dcp$ uncertainty of around $15^\circ$. 
The T2HKK setup allows us to correlate and constrain the systematic errors between the two detectors collecting 
data from the same source, which can increase the sensitivity of the experiment by up to 25\%. 
Such a reduction of the systematic errors is crucial for determining the oscillation parameters 
with greater significance.
\end{abstract} 

\preprint{CTPU-17-10}

\maketitle


\section{Introduction}
\label{sec:intro}
Over the last several decades, experimental tests have confirmed the predictions of the Standard Model
of particle physics to an unprecedented level of precision. The Higgs boson that was discovered at 
the Large Hadron Collider was the last missing piece of the 
Standard Model~\cite{higgs_atlas,higgs_cms}. Various hints have emerged 
for possible new physics beyond the Standard Model at collider experiments. However, the only 
confirmed signal of physics beyond the Standard Model comes from neutrino oscillation 
experiments that point to non-zero neutrino mass, contrary to the Standard Model. 

Neutrino oscillations in the standard three-flavour scenario are parametrized by three mixing angles 
$\theta_{12}$, $\theta_{13}$ and $\theta_{23}$, two mass-squared differences $\dm{21}$ and $\dm{31}$ 
(where $\dm{ij}=m_i^2-m_j^2$) and one CP-violating phase $\dcp$. While $\theta_{12}$, $\theta_{13}$, 
$\dm{21}$ and $|\dm{31}|$ have been measured quite accurately by solar, atmospheric, long-baseline 
and reactor neutrino experiments~\cite{sno,sk,kamland,t2kdisapp,novadisapp,dchooz,dayabay,reno}, 
the global fits of world neutrino data~\cite{global_lisi,global_valle,global_nufit} are inconclusive about the sign of $|\dm{31}|$ or neutrino 
mass hierarchy, the octant of the close-to-maximal mixing angle $\theta_{23}$ and the value of $\dcp$, the 
CP phase\footnote{Recent data from the T2K and \nova\ experiments hint weakly towards a value of $\dcp$ close 
to $-90^\circ$ and normal mass hierarchy~\cite{t2kresults2015,novaresults2016}. 
However, more data is required before 
a statistically significant statement can be made about these parameters~\cite{dcp-90_t2k}.}.
The binary question of whether the neutrino mass hierarchy is normal (NH), i.e. $\dm{31}>0$ 
or inverted (IH), i.e. $\dm{31}<0$ is interesting from the point of view of ruling out possible new 
physics models that favour one hierarchy or the other~\cite{Albright:2006cw}. 
The value of $\theta_{23}$ being exactly $45^\circ$ or very close to it raises the possibility of 
a symmetry in the flavour sector such as $\mu-\tau$ symmetry~\cite{lam}. 
Finally, a precise measurement of $\dcp$ will allow us to observe CP violation in the leptonic sector 
which may explain the matter-antimatter asymmetry of our Universe through the mechanism of 
leptogenesis~\cite{Joshipura:2001ui,Endoh:2002wm}. 

The current generation of long-baseline experiments T2K~\cite{t2k_globes} and \nova~\cite{novareport}, atmospheric 
neutrino experiments Super-Kamiokande (SK)~\cite{sk} and IceCube~\cite{icecube_deepcore}, and
reactor neutrino experiments Double Chooz~\cite{dchooz}, Daya Bay~\cite{dayabay} 
and RENO~\cite{reno} are collecting data which will help to measure these three unknown parameters in 
addition to measuring the other parameters with greater precision. The main hurdle in 
determining the values of the parameters in nature is the existence of degeneracies in this 
multi-dimensional parameter space. If nature has chosen 
favourable combinations of the mass hierarchy and $\dcp$~\cite{novat2k}, 
then T2K and \nova\ themselves will be able to determine the mass hierarchy~\cite{nova_reopt2} 
irrespective of the possible choice of $\dcp$ in such 
scenarios. Meanwhile, a precise measurement of $\dcp$ is strongly correlated 
with a knowledge of the hierarchy and the value of $\theta_{23}$~\cite{uss}. 
However, if nature has chosen unfavourable combinations of the parameters, then 
a statistically significant determination of the unknown parameters will require 
the next generation of experiments. These include the long-baseline experiments 
T2HK~\cite{t2hk} and
DUNE~\cite{dune_cdr_physics} and the atmospheric neutrino experiments Hyper-Kamiokande (HK)~\cite{hkdesignreport}, 
ICAL@INO~\cite{ical} and PINGU~\cite{pingu}. All these experiments will take advantage of matter effects
from neutrinos propagating through the earth. 

The HK proposal envisages augmenting the existing SK detector that has a fiducial 
mass of 22.5 kilotons with two additional Water \v{C}erenkov detector tanks with fiducial mass 187 
kilotons each. This new setup will serve as a detector for both neutrinos from 
the J-PARC beam in Tokai as well 
as atmospheric neutrinos. Recently, an idea that involves placing one tank of fiducial mass 
187 kilotons at a site in Korea in the path of the T2K beam has been proposed~\cite{t2hkk_proposal}. 
(This is actually based on a much older proposal and its subsequent studies~\cite{t2hkk_old1,t2hkk_old2,t2hkk_old3,t2hkk_old4}.) 
This proposal, called Tokai to Hyper-Kamiokande and Korea (T2HKK) will allow the 
observation of neutrinos at the detector in Kamioka as well as neutrinos 
at a far site in Korea that will have experienced matter effects. The advantage of this 
multi-detector setup is that it simultaneously increases the amount of data at 
the T2HK baseline of 295 km and gives access to data at a longer baseline of around 1100 km, 
corresponding to the location of the proposed detector in Korea. 
Since neutrino oscillation probabilities depend strongly on the distance travelled and the 
matter effects experienced by the neutrinos, the addition of information from two different 
baselines is very effective in breaking the parameter 
degeneracies~\cite{twobase1,twobase2,twobase3,twobase4,twobase5,twobase6,lindner,novat2k,lbnoadequate,duneadequate}. 
In particular, the shorter baseline (Tokai to HK) setup is more effective at measuring 
$\dcp$ while the longer one (Tokai to Korea) is better at excluding the wrong hierarchy. 

Yet another long-baseline experiment that is expected to determine the mass hierarchy is 
the upcoming DUNE experiment. This experiment, with a neutrino beam traversing 1300 km 
from Fermilab to SURF will anyway collect neutrino oscillation data with matter 
effects. 
Both HK and DUNE are expected to be deployed in the mid-2020s~\cite{hkdesignreport,dune_timeline}.
This presents the neutrino physics community with two options -- (a) DUNE collects data
at the longer baseline (1300 km) while the beam from Tokai is intercepted by a 374 kiloton 
detector at HK, or (b) DUNE collects data at the longer baseline (1300 km) while the 
beam from Tokai is observed at a 187 kiloton detector at HK and another 187 kiloton detector 
in Korea. In this article, we compare the physics capabilities of these two options 
in measuring $\dcp$ and determining the mass hierarchy\footnote{A recent discussion 
on the complementarity between HK and DUNE can be found in Ref.~\cite{mono_hk_dune}.}. 
We perform this study in 
the standard three-flavour neutrino oscillation framework, i.e. without considering the 
possible existence of sterile neutrino states or non-standard neutrino interactions. 
Some studies of non-standard interactions in the context of T2HKK can be found 
in Refs.~\cite{t2hkk_nsi_mono,t2hkk_nsi_marfatia,t2hkk_syst_mono}. 
All the detectors 
under consideration in this work are capable of 
detecting atmospheric neutrinos in addition to beam neutrinos~\cite{prd2007,ourprl,raj_dune1,raj_dune2}. 
This will significantly 
improve the hierarchy sensitivity of the experiments. However, we do not take these 
data into account in this study. 

This article is organized as follows. In Section~\ref{sec:expt}, we discuss the 
experimental setups that we have considered in this work. Section~\ref{sec:prob} 
discusses the problem of parameter degeneracies and their removal, at the 
level of probabilities. We discuss the results of our numerical simulations in
Section~\ref{sec:res}, before concluding in Section~\ref{sec:sc}.

\section{Experimental setups}
\label{sec:expt}

\subsection{T2HK}

Our simulation of T2HK is based on the description provided in Ref.~\cite{t2hkk_proposal}. 
(See Ref.~\cite{hkdesignreport} for the most recent HK design report.) 
The 
beam from Tokai is a 1.3 MW beam running for ten years (2.5 years in neutrino 
mode and 7.5 years in antineutrino mode). The detector site is located at a 
distance of 295 km from the source,
$2.5^\circ$ off the beam axis. The detector performance is consistent with 
the description in Ref.~\cite{t2hkk_proposal}, and the projected systematic errors are taken 
from the same reference. We use the notation T2HK1/T2HK2 to denote the setup 
with one/two tanks at the HK location, corresponding to a fiducial mass of 
187/374 kilotons. We do not consider staging effects in the deployment of the two detectors. 
In addition to this large HK detector, we also include 
the existing SK detector (22.5 kilotons) whenever we simulate T2HK. 

\subsection{T2Kor}

In this study, the name T2Kor refers to the setup consisting of the source 
at Tokai (the same as for T2HK) and detector in Korea. The description of the 
detector and systematic errors is taken from Ref.~\cite{t2hkk_proposal}. A number of potential 
sites for the detector in Korea have been studied in the literature. Their baselines vary from around 
1000 km to 1200 km, and their off-axis angles vary from around $1^\circ$ to 
$2.5^\circ$. For the purpose of this study, we consider a generic site at a 
distance of 1100 km and off-axis angle of $1.5^\circ$. The neutrino flux 
for this off-axis location is taken from Ref.~\cite{t2hkkfluxweb}. 
The detector mass for this setup is always 187 kilotons.

\subsection{T2HKK}

This is a combination of one detector tank with mass 187 kilotons at the 
HK site, and another (assumed to be identical) detector tank at the Korean 
site. Thus, we have the schematic description 
\[
 \textrm{T2HKK} \equiv \textrm{T2HK1} + \textrm{T2Kor} ~.
\]
Unless stated otherwise, we always correlate the systematics between 
T2HK1 and T2Kor. In our simulations, we do not take into account the delay 
in the deployment of the second detector. 

\subsection{DUNE}

The DUNE experiment that we have considered uses a neutrino beam from Fermilab in the 
1.07 MW -- 80 GeV proton beam configuration, running for 3.5 years each in the 
neutrino and antineutrino modes. The data are collected at a liquid argon 
TPC with fiducial mass 40 kilotons located at SURF, 1300 km away from the source. 
The specifications for this experiment are taken from Ref.~\cite{dune_cdr_physics}. 
Again, we do not consider staging effects in the deployment of the detector.

\section{Probability level discussion and parameter degeneracies}
\label{sec:prob}

The $\numutonue$ oscillation probability $\pmue$ is well suited to measure 
the mass hierarchy and $\dcp$. In terms of the neutrino energy $E$ and 
distance travelled $L$, this probability can be expressed up to 
second order in $s_{13}=\sin\theta_{13}$ and $\alpha=\dm{21}/|\dm{31}|$ 
as~\cite{akhmedov}
\begin{eqnarray}
\pmue & \approx & 4 \ s_{13}^2 \ \sin^2\theta_{23} \ \frac{\sin^2(1-ph\ahat)\Delta}{(1-ph\ahat)^2} \nonumber \\
& + & 2 \ h \ \alpha \ s_{13} \ \sin 2\theta_{12} \ \sin 2\theta_{23} \ \cos(\Delta+ph\dcp) \ \frac{\sin(\ahat \Delta)}{\ahat} \ \frac{\sin(1-ph\ahat)\Delta}{(1-ph\ahat)} ~,
\label{eq:pmue}
\end{eqnarray}
where $\Delta = |\dm{31}|L/4E$ and $\ahat = |2EV/\dm{31}|$ with $V$ denoting 
the Wolfenstein matter potential. We have introduced the binary variables $p = +1(-1)$ 
for neutrinos(antineutrinos) and $h=+1(-1)$ for NH(IH). 

In the limit of 
vacuum oscillations, we have 
\begin{eqnarray}
\pmue (\ahat \to 0) & \approx & 4 \ s_{13}^2 \ \sin^2\theta_{23} \ {\sin^2\Delta} \nonumber \\
& + & 2 \ h \ \alpha \ s_{13} \ \sin 2\theta_{12} \ \sin 2\theta_{23} \ \cos(\Delta+ph\dcp) \ \Delta \ \sin\Delta ~,
\label{eq:pmuevac}
\end{eqnarray}
i.e. the leading order term is independent 
of the hierarchy. In fact, close to the oscillation maximum $\Delta=\pi/2$, the 
second term becomes insensitive to the hierarchy as well. This degeneracy between 
the two hierarchies can be lifted and hierarchy sensitivity restored by increasing 
the matter density that the neutrino travels through, i.e. by going to a 
longer baseline. In the neutrino(antineutrino) mode, the probability 
is higher for NH(IH) than for IH(NH), because of the $(1-ph\ahat)^2$ term 
in the denominator.
Close to the oscillation maximum, the equation for $\pmue$ becomes 
\begin{eqnarray}
\pmue (\Delta=\pi/2) & \approx & 4 \ s_{13}^2 \ \sin^2\theta_{23} \ \frac{\cos^2(\ahat\pi/2)}{(1-ph\ahat)^2} \nonumber \\
& - & 2 \ p \ \alpha \ s_{13} \ \sin 2\theta_{12} \ \sin 2\theta_{23} \ \sin\dcp \ \frac{\sin(\ahat \pi/2)}{\ahat} \ \frac{\cos(\ahat \pi/2)}{(1-ph\ahat)} ~.
\end{eqnarray}
Although the first term increases the probability for NH compared to IH, this 
difference can be offset by the second term if $\dcp=+90^\circ$. Similarly the 
reduced IH probability can be compensated if $\dcp=-90^\circ$. Thus, the 
combinations \{NH, $\dcp=+90^\circ$\} and \{IH, $\dcp=-90^\circ$\} suffer from
the hierarchy-$\dcp$ degeneracy and are unfavourable for measuring these 
parameters. Conversely, determining these parameters is easier for the 
favourable combinations \{NH, $\dcp=-90^\circ$\} and 
\{IH, $\dcp=+90^\circ$\}~\cite{novat2k,nova_reopt2}. 

The $\dcp$ dependence of $\pmue$ is present in the second term of Eq.~\ref{eq:pmue}. 
The measurement of $\dcp$ suffers from the hierarchy-$\dcp$ degeneracy, as 
discussed earlier. However when matter effects are negligible, we see in 
Eq.~\ref{eq:pmuevac} that a clean measurement of $\dcp$ is possible, especially 
near the oscillation maximum. 
(There are additional complications in the measurement of $\dcp$ because of 
the uncertainty in the value of $\theta_{23}$ which affects the first term in
Eq.~\ref{eq:pmuevac}~\cite{uss,minakata2,cpv_long,newton_degen,newton_dune_t23_dcp}. 
The $\dcp$-independent first term acts as a background for the measurement of 
the $\dcp$-dependent second term -- an effect that gets worse with increasing 
$\theta_{23}$.)
Thus, a measurement of the mass hierarchy is 
facilitated by large matter effects but the opposite holds for $\dcp$ measurement. 
In the specific context of T2HKK, it must be noted that the peak of the flux lies close to the 
second oscillation maximum of the T2Kor baseline. Since the variation of 
probability with $\dcp$ given by $\textrm{d}\pmue/\textrm{d}\dcp$ is greater at the 
second oscillation maximum ($\Delta=3\pi/2$) than at the first one ($\Delta=\pi/2$), 
the T2Kor baseline also contributes to the $\dcp$ sensitivity of the T2HKK setup.

DUNE with a baseline of 1300 km will have exceptional sensitivity to the mass 
hierarchy. T2HK on the other hand, because of its relatively short baseline of 
295 km will be able to measure $\dcp$ precisely. If one of the two detector 
tanks of T2HK is moved to a farther location in Korea, it will improve the 
hierarchy sensitivity of the T2HKK setup. On the other hand, since DUNE will 
already collect data with matter effects, will it be preferable to retain 
both detector tanks at the HK site to improve $\dcp$-sensitivity? In the 
next section, we compare the sensitivities of the  
DUNE+T2HK2 and DUNE+T2HKK setups to try and answer this question.

\section{Results}
\label{sec:res}

\begin{figure}[!htb]
\centering
\begin{tabular}{rl}
  \hspace{-0.0in}
  \includegraphics[trim=20 0 10 0,clip,width=0.480\textwidth]{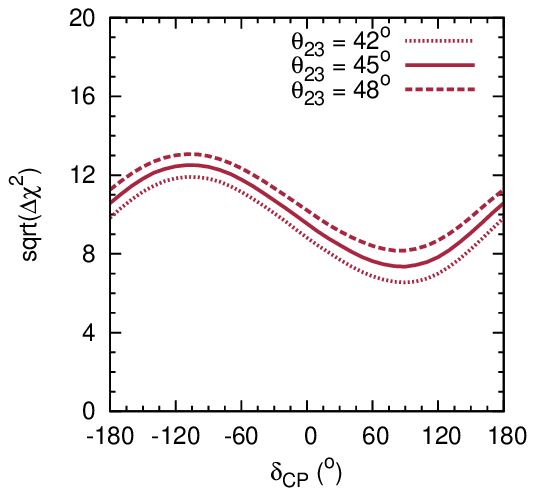} &
  \hspace{-0.5in}
  \includegraphics[trim=20 0 10 0,clip,width=0.480\textwidth]{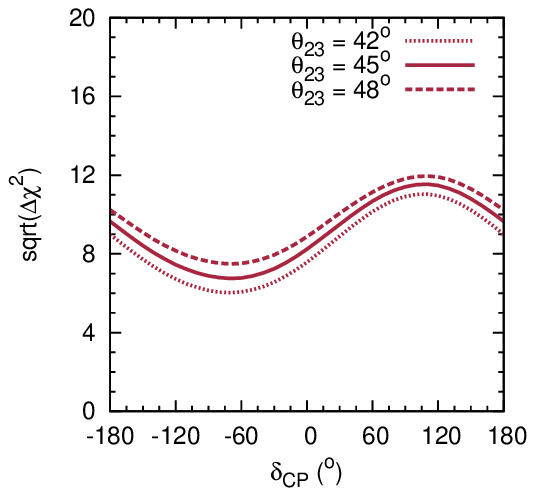} \\
  \hspace{-0.0in}
  \includegraphics[trim=20 0 10 0,clip,width=0.480\textwidth]{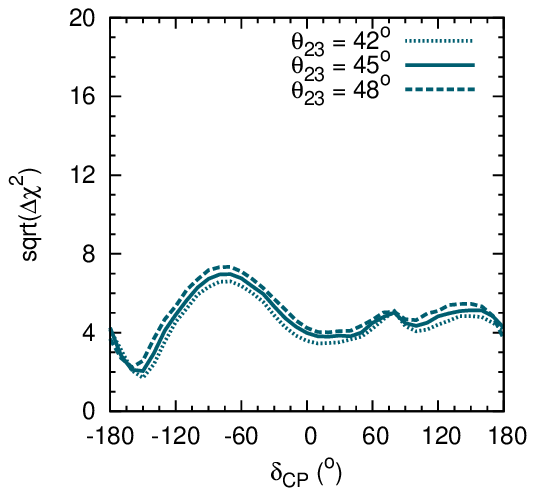} &
  \hspace{-0.5in}
  \includegraphics[trim=20 0 10 0,clip,width=0.480\textwidth]{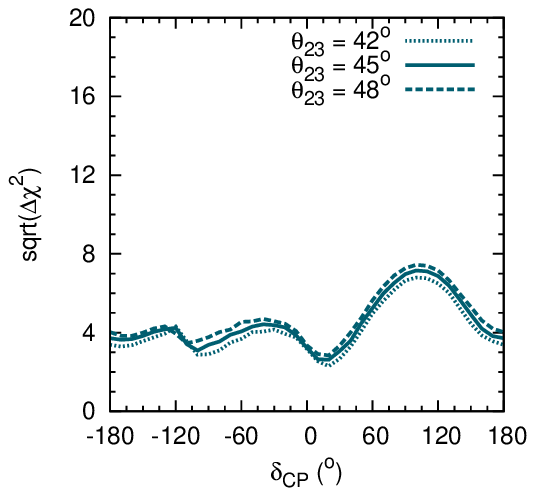} \\
  \hspace{-0.0in}
  \includegraphics[trim=20 0 10 0,clip,width=0.480\textwidth]{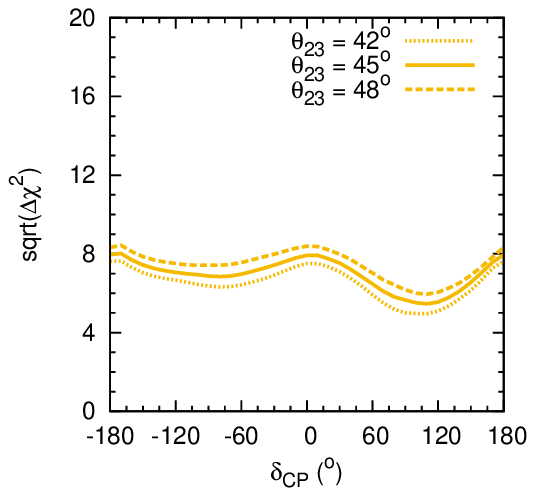} &
  \hspace{-0.5in}
  \includegraphics[trim=20 0 10 0,clip,width=0.480\textwidth]{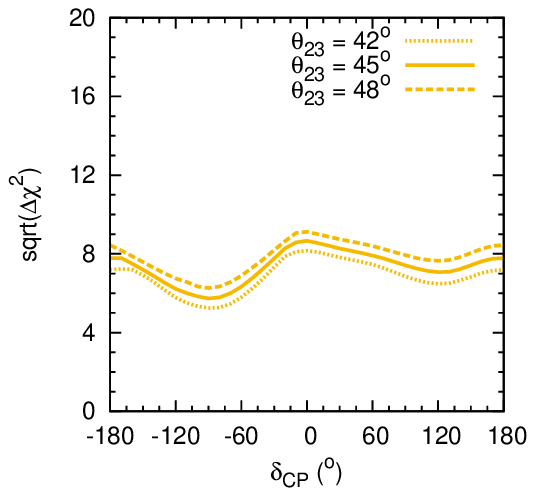} 
 \end{tabular}
\caption{\footnotesize{Hierarchy sensitivity of the experiments DUNE (top row), 
T2HK2 (middle row) and T2Kor (bottom row) as a function of true $\dcp$. The 
curves correspond to three representative values of $\theta_{23}$ and are shown 
for both true hierarchies -- NH (left column) and IH (right column). }}
\label{fig:def_hier}
\end{figure}

\begin{figure}[!htb]
\centering
\begin{tabular}{rl}
  \hspace{-0.0in}
  \includegraphics[trim=20 0 10 0,clip,width=0.48\textwidth]{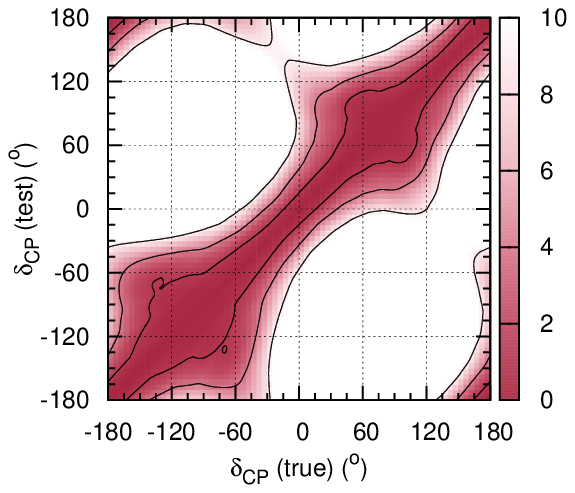} &
  \hspace{-0.5in}
  \includegraphics[trim=20 0 10 0,clip,width=0.48\textwidth]{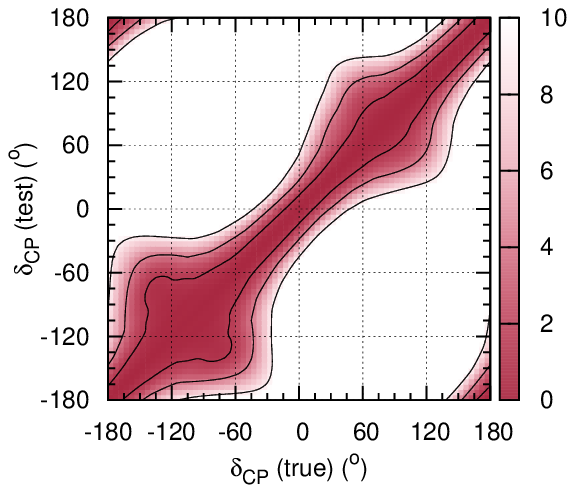} \\
  \hspace{-0.0in}
  \includegraphics[trim=20 0 10 0,clip,width=0.48\textwidth]{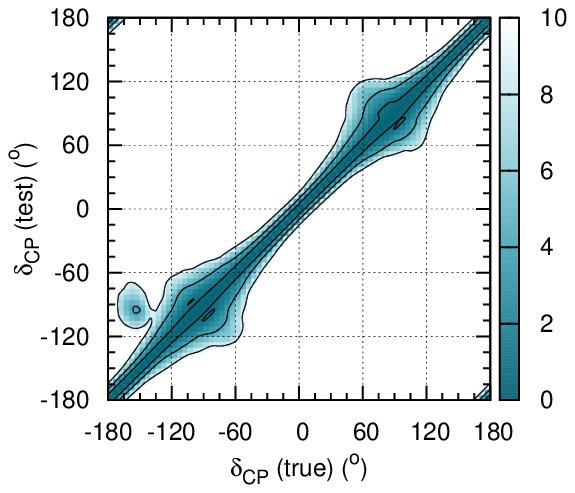} &
  \hspace{-0.5in}
  \includegraphics[trim=20 0 10 0,clip,width=0.48\textwidth]{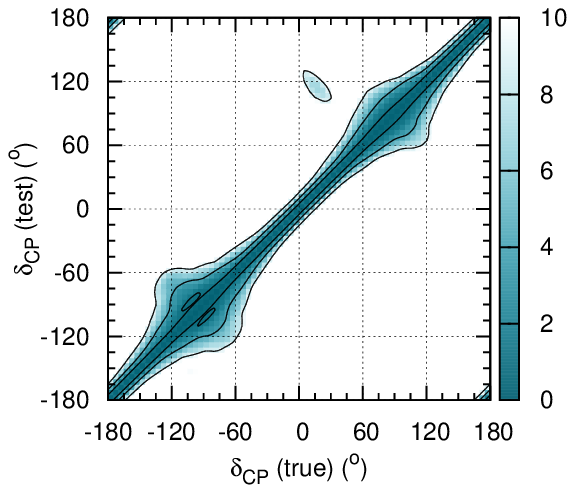} \\
  \hspace{-0.0in}
  \includegraphics[trim=20 0 10 0,clip,width=0.48\textwidth]{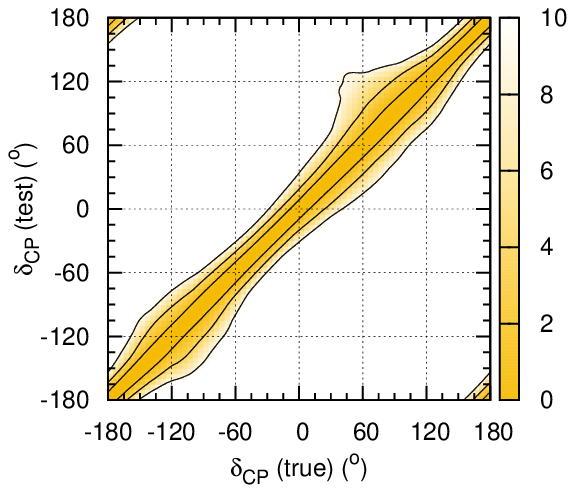} &
  \hspace{-0.5in}
  \includegraphics[trim=20 0 10 0,clip,width=0.48\textwidth]{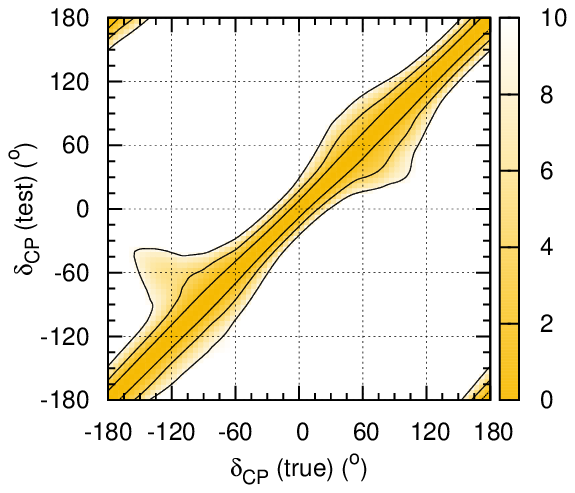} 
 \end{tabular}
\caption{\footnotesize{The allowed range of test $\dcp$ for the experiments DUNE (top row), 
T2HK2 (middle row) and T2Kor (bottom row) as a function of true $\dcp$, for 
true $\theta_{23}=45^\circ$ and both true hierarchies -- NH (left column) and 
IH (right column). The coloured shading along the z-axis represent $\Delta\chi^2$ values, and the
contours correspond to $\sqrt{\Delta\chi^2}=1,2,3$.}}
\label{fig:def_cpallowed}
\end{figure}

\begin{figure}[!htb]
\centering
\begin{tabular}{rl}
  \hspace{-0.0in}
  \includegraphics[trim=20 0 10 0,clip,width=0.47\textwidth]{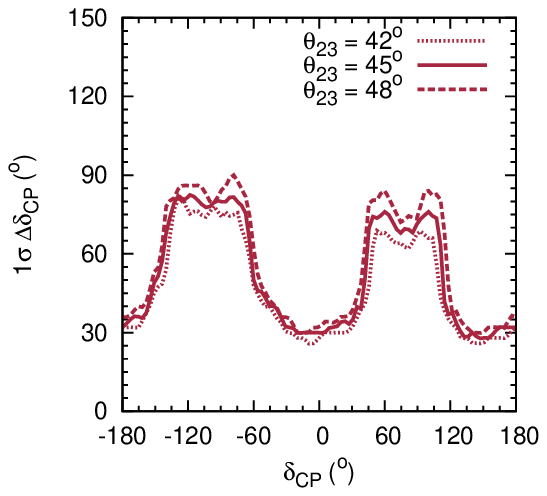} &
  \hspace{-0.5in}
  \includegraphics[trim=20 0 10 0,clip,width=0.47\textwidth]{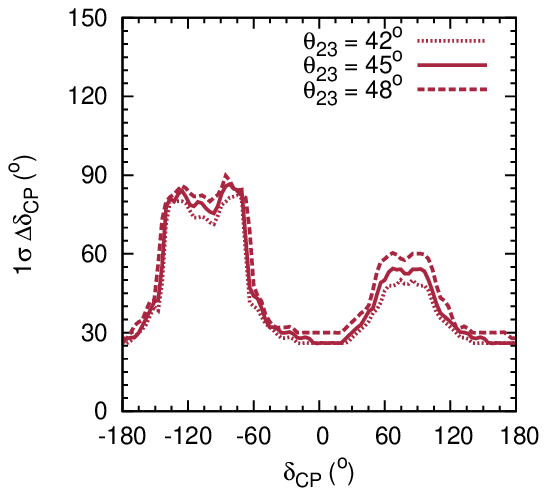} \\
  \hspace{-0.0in}
  \includegraphics[trim=20 0 10 0,clip,width=0.47\textwidth]{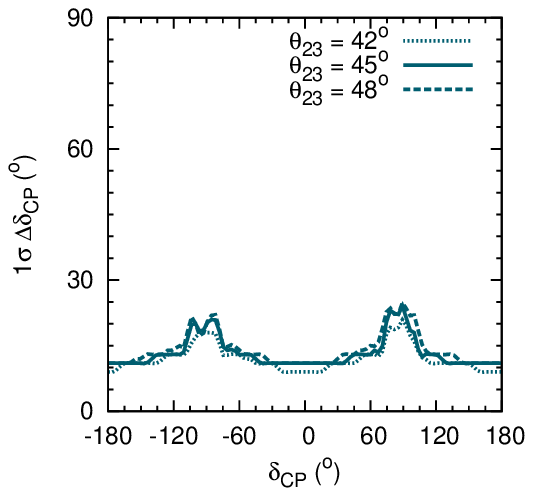} &
  \hspace{-0.5in}
  \includegraphics[trim=20 0 10 0,clip,width=0.47\textwidth]{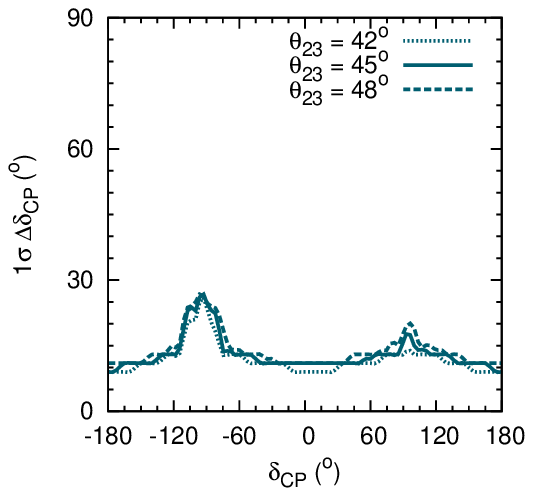} \\
  \hspace{-0.0in}
  \includegraphics[trim=20 0 10 0,clip,width=0.47\textwidth]{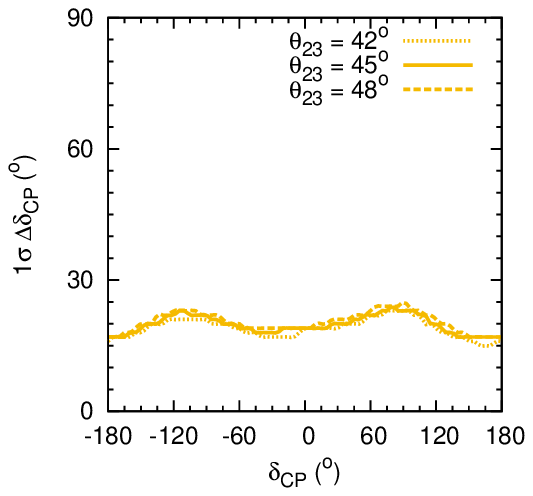} &
  \hspace{-0.5in}
  \includegraphics[trim=20 0 10 0,clip,width=0.47\textwidth]{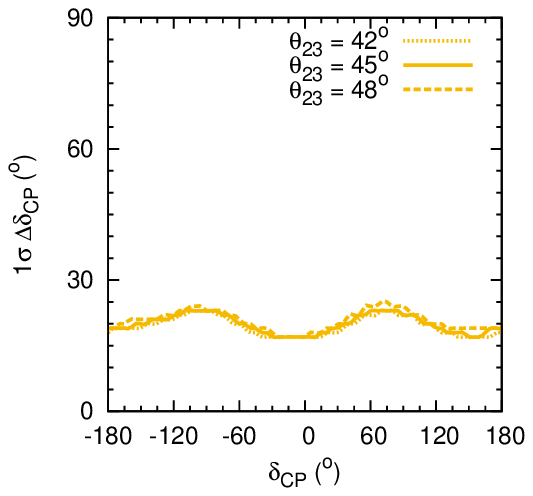} 
 \end{tabular}
\caption{\footnotesize{The $1\sigma$ ($\sqrt{\Delta\chi^2}=1$) uncertainty in the allowed $\dcp$ range 
for the experiments DUNE (top row), 
T2HK2 (middle row) and T2Kor (bottom row) as a function of true $\dcp$. The 
curves correspond to three representative values of $\theta_{23}$ and are shown 
for both true hierarchies -- NH (left column) and IH (right column). The uncertainties 
are calculated as $360^\circ \times$ the fraction of allowed $\dcp$ values and correspond to 
twice the usual error.}}
\label{fig:def_errorcp}
\end{figure}

We now discuss the results of our numerical simulations. The experimental 
setups descibed in Section~\ref{sec:expt} are simulated using the GLoBES 
package~\cite{globes,globes2} along with its auxiliary files~\cite{globes_xsec,globes_xsec2}. 
For a given set of `true' and `test' oscillation parameters $\vec{p}_\textrm{true}$ 
and $\vec{p}_\textrm{test}$, the binned true and test event rates 
$N_i^\textrm{true}$ and $N_i^\textrm{test}$ are simulated using GLoBES. 
The test rates are modified $N_i^\textrm{test} \to N_i^\textrm{test}(\xi)$ 
to take systematic effects into account through the nuisance parameter(s) $\xi$. 
We then perform a $\chi^2$ analysis using 
the method of pulls~\cite{pulls,pulls2,pulls3} 
\begin{equation}
\chi^2 (\vec{p}_\textrm{true},\vec{p}_\textrm{test}) = \min_\xi \left[ \left( \sum_{i\in\textrm{bins}} \frac{(N_i^\textrm{true}-N_i^\textrm{test}(\xi))^2}{N_i^\textrm{true}} \right) + \frac{\xi^2}{\sigma_\xi^2} \right] ~,
\end{equation}
where $\sigma_\xi$ is the $1\sigma$ systematic error corresponding to 
the experiment in question. 
In simulating the T2HKK setup, the systematics between the `constituent 
experiments' T2HK1 and T2Kor are correlated, unless specified otherwise. 
In this correlated case, the $\chi^2$ is calculated as
\begin{equation}
\chi^2 (\vec{p}_\textrm{true},\vec{p}_\textrm{test}) = \min_\xi \left[ \left( \sum_{e\in\textrm{expt}} \sum_{i\in\textrm{bins}} \frac{(N_{e,i}^\textrm{true}-N_{e,i}^\textrm{test}(\xi))^2}{N_{e,i}^\textrm{true}} \right) + \frac{\xi^2}{\sigma_\xi^2} \right] ~.
\label{eq:chisqcorrel}
\end{equation}

In our simulations, we marginalize over the $3\sigma$ ranges of $\theta_{13}$, 
$\theta_{23}$, $\dm{31}$ and $\dcp$ given by the global fit~\cite{global_nufit}. 
We include a Gaussian prior on $\sin^2 2\theta_{13}$ with an error 
of $0.005$, to account for the reactor neutrino constraint on 
this parameter. While marginalizing over $\dm{31}$ and $\theta_{23}$ 
in discrete steps, we find it useful to include the three-flavour 
corrections~\cite{dm31_defn,dm31_defn2,th23_defn} in order to avoid spurious contributions 
to the sensitivity. 
We do not assume that the mass hierarchy 
is known, i.e. we allow the test value of $\dm{31}$ to be both positive 
and negative and choose the minimum of the two.

\subsection{Typical sensitivity of individual setups}

We first show the hierarchy exclusion sensitivity of DUNE, T2HK2 and T2Kor in 
Fig.~\ref{fig:def_hier}. The results are shown as a function of the true value 
of $\dcp$ for both hierarchies and three typical values of true 
$\theta_{23}$. The sensitivity is represented by the quantity $\sqrt{\Delta\chi^2}$. 
(If the conditions required by Wilks' theorem are satisfied, the computed 
$\Delta\chi^2$ follows the $\chi^2$-distribution and $n=\sqrt{\Delta\chi^2}$  
simply indicates the $n\sigma$ confidence level. However, due to the 
fact that mass hierarchy is a discrete binary parameter, the relation 
between $\sqrt{\Delta\chi^2}$ and the statistical confidence levels 
is not trivial. We refer the reader to Refs.~\cite{evslin_stat,blennow_stat1,blennow_stat2} for detailed 
discussions on this matter.) 
Irrespective of the value of $\dcp$ or hierarchy, both DUNE and T2Kor 
are capable of excluding the wrong hierarchy with $\sqrt{\Delta\chi^2}>5$.

As expected based on the discussion in Section~\ref{sec:prob}, 
the hierarchy sensitivity is greater for the favourable combinations 
of hierarchy and $\dcp$. We also see that the hierarchy sensitivity of DUNE 
is the highest while that of T2HK2 is the lowest, due to matter effects. 
Finally we observe that hierarchy sensitivity increases with $\theta_{23}$, 
owing to the leading order term in the oscillation probability. 

In Fig.~\ref{fig:def_cpallowed}, we show the values of test $\dcp$ that are 
consistent with a given value of true $\dcp$. The colour coding along the z-axis 
in these contour plots represents the value of $\Delta\chi^2$, while 
the equi-precision contours have been draw for $\sqrt{\Delta\chi^2}=1,2,3$. 
The plots shown here are for 
true $\theta_{23}=45^\circ$. As expected, the allowed 
values lie close to the $\textrm{true } \dcp = \textrm{test } \dcp$ line. 
Here, we find that T2HK2 has the best precision in measuring $\dcp$, 
followed by T2Kor and finally DUNE because of matter 
effects. The sensitivity is seen to be best around $\dcp=0,\pm180^\circ$ 
and worst around $\pm 90^\circ$. This is because the precision in 
$\dcp$ is proportional to $\textrm{d}\pmue/\textrm{d}\dcp$ which goes as 
$\cos\dcp$ near the oscillation maximum. 

A more intuitive way of seeing these results is to plot the 
uncertainty in the measurement of $\dcp$ (at the $\sqrt{\Delta\chi^2}=1$ level) as a function of its true value. This 
is shown in Fig.~\ref{fig:def_errorcp}. The uncertainty plotted is simply 
$360^\circ \times$ the fraction of allowed $\dcp$ values seen in Fig.~\ref{fig:def_cpallowed}. 
In most cases this corresponds to twice the `usual' 
error -- an important point while comparing these results with the 
collaboration reports Ref.~\cite{dune_cdr_physics,t2hkk_proposal}. (Exceptions are when the allowed range is not symmetric around 
the true value. See for example the middle left panel of Fig.~\ref{fig:def_cpallowed} 
near true $\dcp=-160^\circ$.) We find that T2HK2 has the best precision in $\dcp$ 
with an uncertainty of $10^\circ-15^\circ$ for most values of $\theta_{23}$ and $\dcp$, 
except when $\dcp$ is close to $\pm 90^\circ$. Note that the mass hierarchy is a free 
parameter in our analysis. The precision in $\dcp$ is found to improve 
if the mass hierarchy is known.

\subsection{Effect of systematics}

\begin{figure}[htb]
\centering
\begin{tabular}{rl}
  \hspace{-0.0in}
  \includegraphics[trim=20 0 10 0,clip,width=0.48\textwidth]{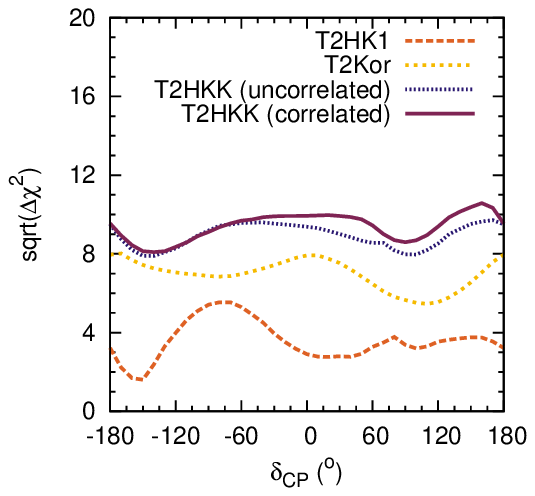} &
  \hspace{-0.5in}
  \includegraphics[trim=20 0 10 0,clip,width=0.48\textwidth]{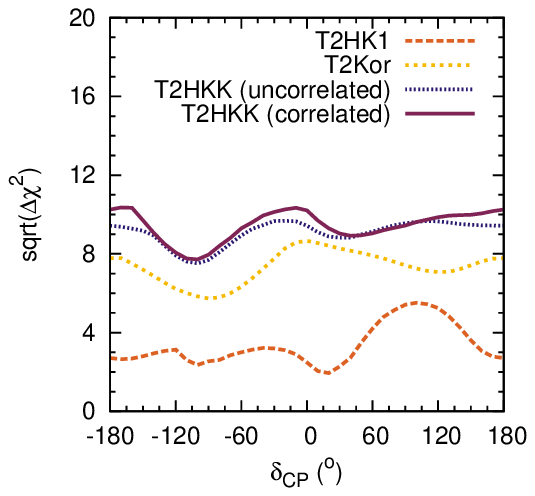} 
\end{tabular}
\caption{\footnotesize{Hierarchy sensitivity as a function of true $\dcp$ 
for the experiments T2HK1, 
T2Kor, and their combination T2HKK considering both cases -- correlated and 
uncorrelated systematics. The 
curves correspond to $\theta_{23}=45^\circ$ and are shown 
for both true hierarchies -- NH (left) and IH (right).}}
\label{fig:systcorrel}
\end{figure}

The two constituent long-baseline setups of T2HKK -- T2HK1 and T2Kor 
share a common beam source. The detector technology and size, and hence 
the detector capabilities are also expected to be the same. Therefore, 
there is a strong correlation between the systematic effects experienced 
by these two setups. In this subsection, 
we consider two cases for the combined T2HKK setup -- one in which the 
systematic errors between the two constituents are uncorrelated, and the 
other where they are correlated. In the uncorrelated case, the 
total $\chi^2$ is simply the sum of the $\chi^2$ values from the 
pull analyses of the individual experiments\footnote{Ref.~\cite{t2hkk_syst_mono} 
discusses the effect of uncorrelated systematics 
in the context of T2HKK.}.
The calculation in 
the correlated case follows the method described by Eq.~\ref{eq:chisqcorrel}. 
In Fig.~\ref{fig:systcorrel} we show the result of our computations 
for the hierarchy exclusion capability. For completeness, we have also 
showed the hierarchy sensitivity of the individual setups T2HK1 and T2Kor. We find an increase 
in $\Delta\chi^2$ of up to 25\% when the systematic errors are correlated between these 
experiments. The analysis performed here is very simplistic 
in its assumptions about the systematic effects, making use of a few effective 
nuisance parameters. In addition, we assume complete correlation between the 
systematics at the two constitutent setups. Therefore, our results 
(for hierarchy sensitivity as well as $\dcp$ precision) are 
slightly more optimistic than the ones in Ref.~\cite{t2hkk_proposal}.  
Throughout the rest of this 
study, we will perform our analysis using correlated systematics.

\subsection{Performance of T2HKK}

\begin{figure}[!htb]
\centering
\begin{tabular}{rl}
  \hspace{-0.0in}
  \includegraphics[trim=20 0 10 0,clip,width=0.48\textwidth]{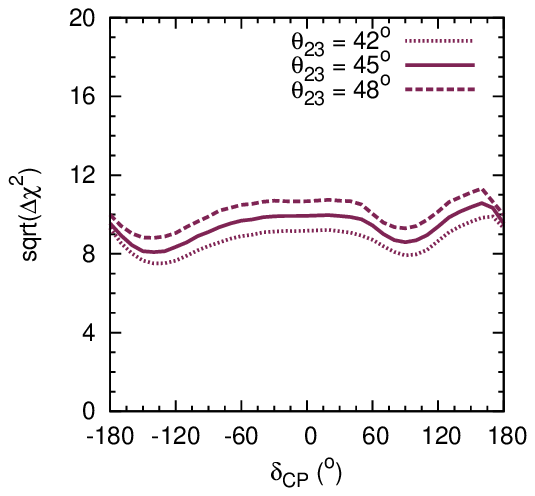} &
  \hspace{-0.5in}
  \includegraphics[trim=20 0 10 0,clip,width=0.48\textwidth]{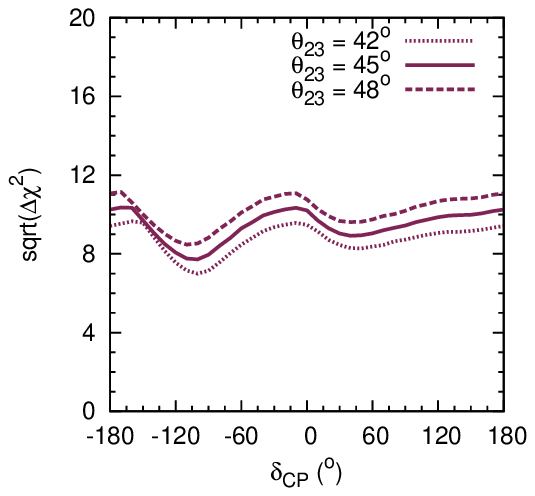} \\
  \hspace{-0.0in}
  \includegraphics[trim=20 0 10 0,clip,width=0.48\textwidth]{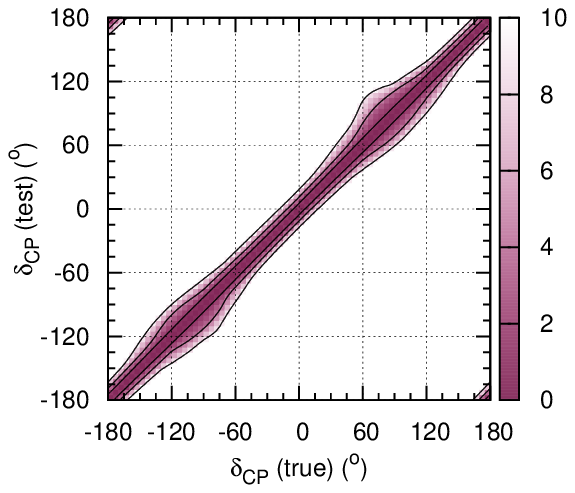} &
  \hspace{-0.5in}
  \includegraphics[trim=20 0 10 0,clip,width=0.48\textwidth]{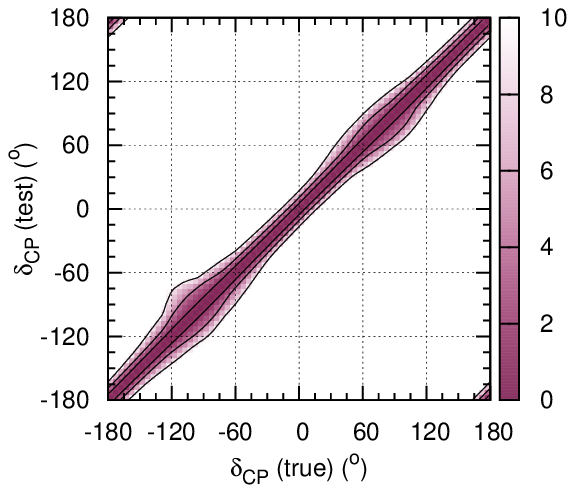} \\
  \hspace{-0.0in}
  \includegraphics[trim=20 0 10 0,clip,width=0.48\textwidth]{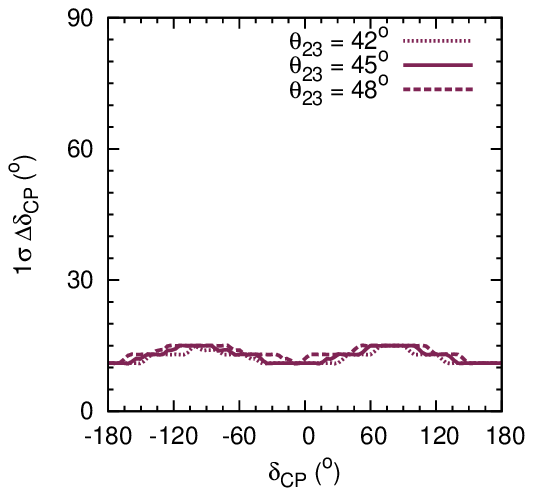} &
  \hspace{-0.5in}
  \includegraphics[trim=20 0 10 0,clip,width=0.48\textwidth]{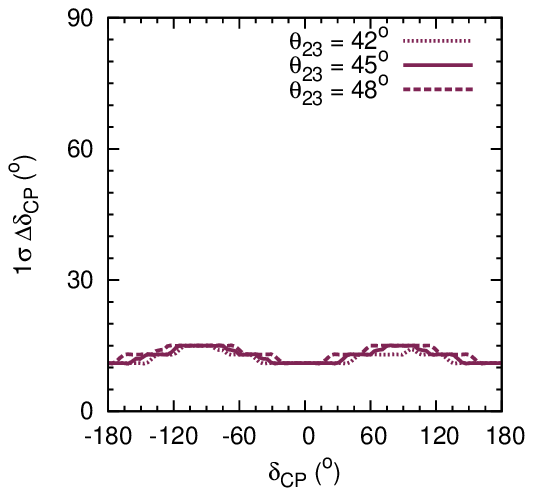} 
\end{tabular}
\caption{\footnotesize{Top row: Hierarchy sensitivity of T2HKK 
as a function of true $\dcp$. Middle row: The allowed range of test $\dcp$ for 
T2HKK, for $\theta_{23}=45^\circ$. Bottom row: The $1\sigma$ ($\sqrt{\Delta\chi^2}=1$) uncertainty 
of the allowed $\dcp$ range for T2HKK. 
All the plots are shown 
for both true hierarchies -- NH (left) and IH (right).}}
\label{fig:t2hkk}
\end{figure}

Figure~\ref{fig:t2hkk} shows the performance of the T2HKK setup in 
excluding the wrong mass hierarchy and in measuring $\dcp$ precisely. 
As the top panel shows, the wrong hierarchy can be excluded at 
$\sqrt{\Delta\chi^2} \gsim 7$, irrespective of the hierarchy and values of $\theta_{23}$ 
and $\dcp$. Comparing with Fig.~\ref{fig:def_hier}, we see that the 
performance is not as good as DUNE for favourable values of $\dcp$ because 
DUNE has more matter effects. However, it is definitely better than DUNE for 
unfavourable values of $\dcp$ because the shorter baseline data helps 
to lift the hierarchy-$\dcp$ degeneracy by constraining $\dcp$. 
The allowed ranges of $\dcp$ seen in the middle row 
(shown for $\theta_{23}=45^\circ$) are much smaller than those of 
the T2Kor setup shown in Fig.~\ref{fig:def_cpallowed}, thanks 
to data from the shorter baseline that favours CP-measurement. The 
$\sqrt{\Delta\chi^2}=1$ allowed range is seen to run almost parallel to the 
diagonal, giving a precision of $10^\circ-15^\circ$, which is reflected in 
the lower panel. This corresponds (roughly) to an error of 
around $5^\circ-7.5^\circ$ in the measurement of $\dcp$.

\subsection{Comparison of setups}

\begin{figure}[!htb]
\centering
\begin{tabular}{rl}
  \hspace{-0.0in}
  \includegraphics[trim=20 0 10 0,clip,width=0.47\textwidth]{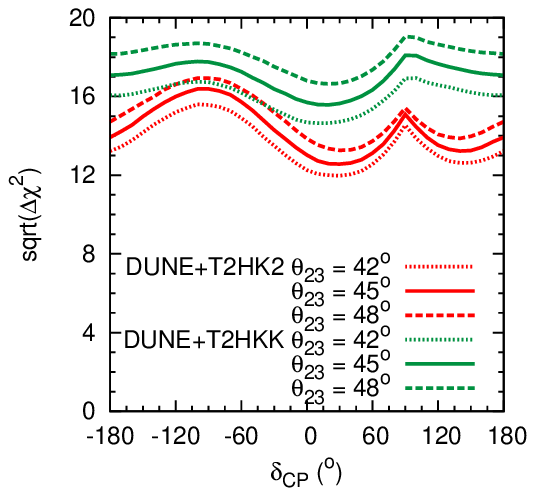} &
  \hspace{-0.5in}
  \includegraphics[trim=20 0 10 0,clip,width=0.47\textwidth]{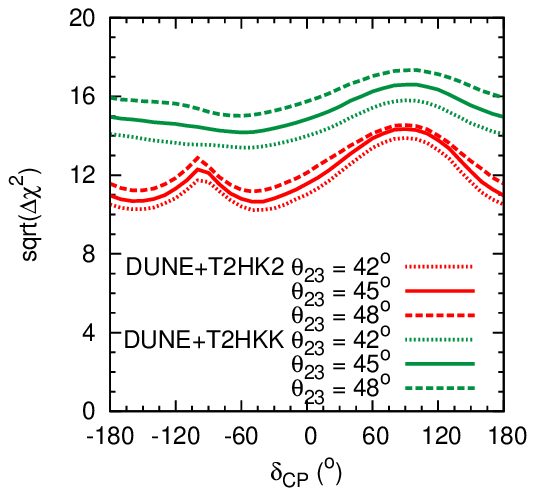} \\
  \hspace{-0.0in}
  \includegraphics[trim=20 0 10 0,clip,width=0.47\textwidth]{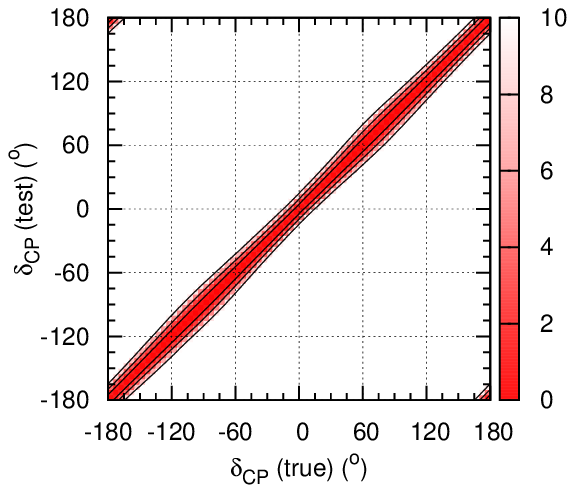} &
  \hspace{-0.5in}
  \includegraphics[trim=20 0 10 0,clip,width=0.47\textwidth]{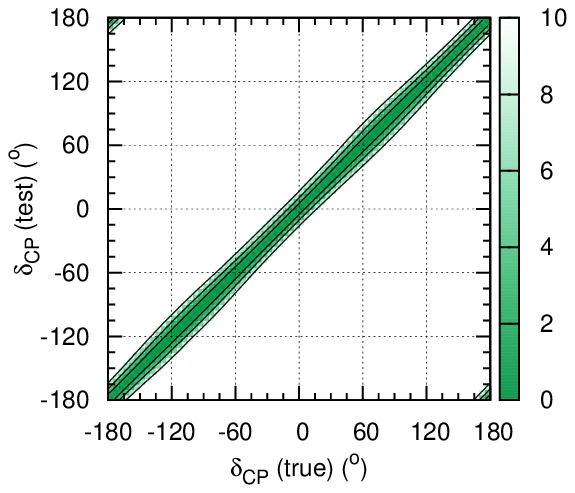} \\
  \hspace{-0.0in}
  \includegraphics[trim=20 0 10 0,clip,width=0.47\textwidth]{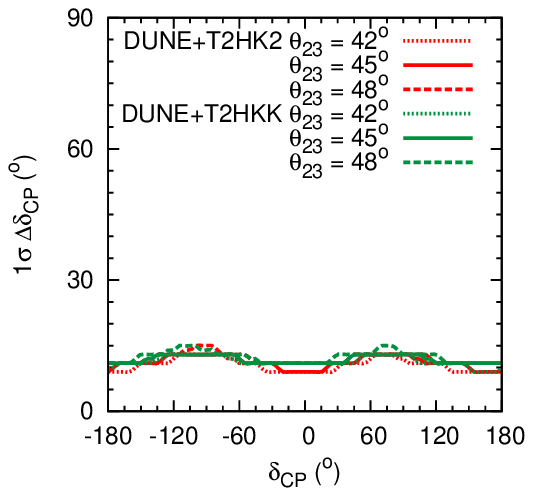} &
  \hspace{-0.5in}
  \includegraphics[trim=20 0 10 0,clip,width=0.47\textwidth]{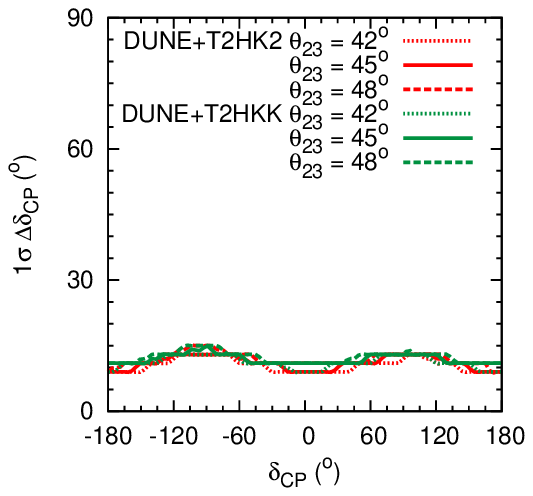} 
\end{tabular}
\caption{\footnotesize{Physics reach for the combined analysis of 
DUNE+T2HK2 and DUNE+T2HKK. 
Top row: Hierarchy sensitivity of DUNE+T2HK2 and DUNE+T2HKK
as a function of true $\dcp$. The plots are shown 
for both true hierarchies -- NH (left) and IH (right). Middle row: The allowed range of test $\dcp$ 
for DUNE+T2HK2 (left) and DUNE+T2HKK (right), for 
true $\theta_{23}=45^\circ$ and true NH. Bottom row: The $1\sigma$ ($\sqrt{\Delta\chi^2}=1$) uncertainty 
of the allowed $\dcp$ range for DUNE+T2HK2 and DUNE+T2HKK. 
The plots are shown 
for both true hierarchies -- NH (left) and IH (right).}}
\label{fig:finalcompare}
\end{figure}

Finally, we compare the capabilities of the T2HK2 and T2HKK setups 
in conjunction with DUNE. Since DUNE will anyway collect data at a 
baseline of 1300 km, these results tell us whether there is any 
advantage in installing one of the two HK detector tanks in Korea. 
The top row of Fig.~\ref{fig:finalcompare} shows the hierarchy discriminating ability of the setups in 
question as a function of $\dcp$, for three representative values 
of $\theta_{23}$. For all the parameter values under 
consideration, the wrong hierarchy can be excluded with $\sqrt{\Delta\chi^2}\gsim10(13)$ 
by DUNE+T2HK2 (DUNE+T2HKK). In other words, 
the combination DUNE+T2HKK outperforms DUNE+T2HK2 by a wide margin. 
This is to be expected since T2HKK has more matter effects which 
help in distinguishing the two hierarchies. The middle row shows the 
allowed $\dcp$ regions for the two setups for NH and $\theta_{23}=45^\circ$. 
Thanks to data from T2HK and the high hierarchy-discriminating ability of 
the longer baselines, the precision in $\dcp$ is high. The uncertainty
in $\dcp$ is plotted in the bottom row. DUNE+T2HKK has a 
precision between $10^\circ$ and $15^\circ$, which is the same precision as T2HKK. 
On the other hand, DUNE+T2HK2 has better 
precision close to the CP-conserving values of $\dcp=0,180^\circ$, 
which is an improvement over the precision of T2HK2 alone. 
Close to the current best-fit 
value of $-90^\circ$, the performance of both setups is comparable. 

\begin{figure}[!htb]
 \centering
 \begin{tabular}{rl}
   \hspace{-0.0in}
   \includegraphics[trim=20 0 10 0,clip,width=0.5\textwidth]{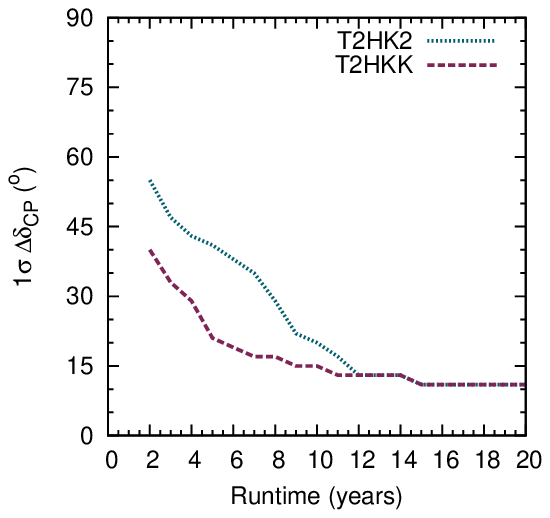} 
   \hspace{-0.5in}
   \includegraphics[trim=20 0 10 0,clip,width=0.5\textwidth]{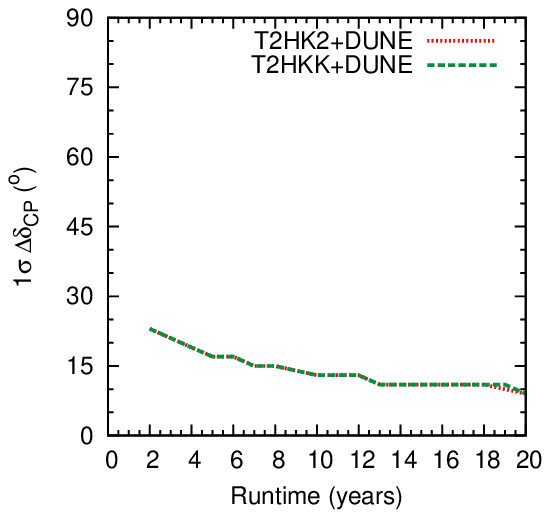} 
 \end{tabular}
 \caption{\footnotesize{Left panel: The $1\sigma$ ($\sqrt{\Delta\chi^2}=1$) uncertainty in $\dcp$ for the 
 experiments T2HK2 and T2HKK as a function of the runtime of the experiment. The 
 curves correspond to $\theta_{23}=45^\circ$, $\dcp=-90^\circ$ and NH. Right panel: Same as left panel, 
 but in conjunction with full exposure of DUNE.}}
 \label{fig:exposat}
 \end{figure}

 In Fig.~\ref{fig:exposat}, we show the uncertainty in the measurement 
of $\dcp$ with T2HK2 and T2HKK as a function of the runtime of the experiment. 
The plots have been generated assuming true values $\theta_{23}=45^\circ$, 
$\dcp=-90^\circ$ and NH.
Here, 10 years corresponds to the benchmark exposure that has been 
considered throughout this work. Initially as the exposure of the 
experiment is increased, we see from the left panel that the error in the measurement of $\dcp$ is 
smaller for T2HKK than for T2HK2. This is because T2HKK is capable of 
breaking the hierarchy-$\dcp$ degeneracy because of its long-baseline 
component. T2HK2 requires around 12 years to achieve the same precision as T2HKK.
Further, we see that increasing the 
runtime beyond 12 years only marginally increases the 
$\dcp$-precision, i.e. there is a saturation of the precision. 
We have checked that in the ideal case of zero systematics, 
the $\dcp$-precision with T2HK2(T2HKK) can be reduced by around $5^\circ$($2^\circ$) 
over a runtime of 10 years. 
This emphasizes 
the role of controlling systematic errors in order to measure 
$\dcp$ with very high precision. With the T2HKK setup, we have a 
unique configuration with two far detectors in addition to a near detector collecting data using 
the same source. Improved measurements of the flux and cross-section using 
this multi-detector setup will 
allow us to severely constrain the 
systematic effects, contributing towards a more precise measurement of $\dcp$.
The curves in the right panel show the corresponding sensitivity after including 
the full exposure of DUNE. 
The combined sensitivity is better than the ones in the left panel, 
as expected. In addition, we see that both combinations perform equally well. 
Note that the question of which combination is better will depend on the true 
value of $\dcp$, $\theta_{23}$ and mass hierarchy in nature, as seen from the bottom 
panels of Fig.~\ref{fig:finalcompare}. The difference between the $\dcp$-uncertainty 
of both combinations is always less than $3^\circ$.

\section{Summary and conclusions}
\label{sec:sc}

The recent T2HKK proposal envisages placing one of the two detector 
tanks of HK at a distant site in Korea in the path of the J-PARC 
neutrino beam. This will allow the observation of neutrinos after 
propagating for a distance of around 1100 km, in addition to the 
data being collected by the first detector tank at the original 
HK site in Japan. The combined setup will be capable of observing
neutrino oscillations in the presence (absence) of matter effects 
through the detector in Korea (Japan). 

The measurement of the neutrino mass hierarchy and $\dcp$ are two of 
the outstanding problems in neutrino physics today. Efforts towards 
measuring them at current facilities are impeded by the 
hierarchy-$\dcp$ degeneracy. This degeneracy can be lifted and the 
hierarchy can be measured in the presence of substantial matter effects. 
On the other hand, a precise measurement of $\dcp$ requires high 
statistics and small matter effects. Thus, the T2HKK setup is
well suited to simultaneously measure both these unknowns. 

Meanwhile, the DUNE experiment will collect neutrino oscillation 
data with a baseline of 1300 km, with substantial matter effects. 
Naturally, such a setup has been shown to have very good 
sensitivity to the mass hierarchy. In light of the fact that 
DUNE will resolve the issue of the mass hierarchy, it is worth 
investigating whether the second HK tank should be placed in 
Korea (giving more matter effects and hierarchy sensitivity)
or at the HK site itself (giving high statistics with small 
matter effects, to enhance CP sensitivity). In this work, we have 
studied the combined capability of DUNE+T2HK2 and DUNE+T2HKK 
to address this issue. 

First, we have presented the capabilities of the individual 
setups towards determining the mass hierarchy and measuring 
$\dcp$. We find the physics reach of these setups in accordance 
with our physical understanding of the physics of neutrino 
oscillations, matter effects and parameter degeneracies. 
We then combine the T2HK1 and T2Kor setups to test the capability 
of T2HKK. Through a simplistic treatment of the systematic effects, 
we show that performing a correlated systematic analysis increases 
the sensitivity of the experiment by up to 25\%. We find that this setup is 
capable of excluding the wrong hierarchy with 
$\sqrt{\Delta\chi^2}>7$, and measuring $\dcp$ with an 
uncertainty of around $10^\circ-15^\circ$.

Finally, we conduct a comparison of the setups DUNE+T2HK2 and 
DUNE+T2HKK. As far as hierarchy discrimination is concerned, 
DUNE+T2HKK with an exclusion sensitivity of $\sqrt{\Delta\chi^2}\gsim13$ 
outperforms DUNE+T2HK2 purely because of more
matter effects. Over a large part of the parameter 
space including around the current best-fit point, the 
two setups perform equally well, with a $\dcp$ uncertainty of 
around $10^\circ-15^\circ$. 
The precision can be improved by $2^\circ-5^\circ$ by reducing the systematics 
(which is possible with a multi-detector setup like T2HKK)
and collecting more data.
While an improvement from $15^\circ$ to $10^\circ$ in the precision 
of $\dcp$ may not seem like much, it is significant if $\dcp$ truly 
lies close to the maximally CP-violating value of $-90^\circ$ which 
could signal a new symmetry of nature.
In conclusion, both setups offer the possibility of 
measuring the unknown parameters to very good precision. 
While the setup with T2HKK is better at determining the 
mass hierarchy, the setup involving T2HK2 gives the same or better 
$\dcp$ precision, depending on the parameter values chosen. 
For the current best-fit values, the capabilities of the 
two setups are comparable.

\section*{Acknowledgements}
The author would like to thank Monojit Ghosh, Kaoru Hagiwara and Hye-Sung Lee for useful discussions. 
This work was supported by IBS under the project code IBS-R018-D1.

\bibliographystyle{JHEP} 
\bibliography{t2hkk}

\end{document}